\documentclass[aps, prb, twocolumn]{revtex4-2}
\usepackage{graphicx, subfigure,braket}
\usepackage{amsmath,amssymb}
\usepackage{color}
\usepackage{natbib, hyperref}
\usepackage[thinlines]{easytable}
\usepackage{comment}
\DeclareMathOperator{\sgn}{sgn}
\begin{document}

\title{Enhanced local addressability of a spin array with local exchange pulses and global microwave driving}

\author{Anoosha Fayyaz}
\email{afayyaz1@umbc.edu}

\author{J.~P.~Kestner}%
\affiliation{Department of Physics, University of Maryland, Baltimore County, Baltimore, MD 21250, USA}

\begin{abstract} 
We theoretically propose a strategy to address an individual spin in a large array of spin qubits with a random distribution of g-factors by employing a combination of single-qubit and {\sc swap} gates facilitated by a global microwave field and local exchange pulses. Consequently, only the target qubit undergoes the desired operation and all other qubits return to their original states, even qubits that share the same Larmor frequency as the target. Gate fidelities above 99\% can thus be maintained for arrays containing tens of qubits.

\end{abstract}

\maketitle

\section{Introduction}

Quantum computers are intrinsically capable of outperforming all known classical algorithms for prime factorization \cite{ekert1996quantum}, and there is an ongoing search for applications to more general tasks in machine learning \cite{biamonte2017quantum, benedetti2017quantum}, deep learning \cite{wiebe2014quantum, verdon2018universal, wilson2021quantum}, etc. Such a prospect motivates intense industrial interest in quantum computing, though the excitement is tempered by uncertainty as to whether there is a practical quantum advantage in such applications \cite{Ciliberto2018,Tang2023}.  
To protect qubits against noise and build reliable quantum computers, we need quantum error correction \cite{gottesman2010introduction}. However, full use of error-correcting codes is expected to require millions of physical qubits \cite{preskill2018quantum}, which makes scalability an inevitable step in the development of quantum computing technologies. The spin states of electrons in quantum dots, naturally forming two-level systems, are a fantastic candidate for the realization of qubits owing to their long coherence times \cite{veldhorst2014addressable}, vast nanofabrication industry at their disposal, and operational temperatures well above 1K \cite{yang2020operation}. An indicator of the success of the spin qubit approach is the demonstration of high-fidelity single-qubit \cite{petit2020universal, yoneda2018quantum, yang2020operation, noiri2022fast, mills2022high} and two-qubit \cite{Xue2022, noiri2022fast, mills2022two} gates in silicon quantum devices with fidelities above the surface code error-correction threshold \cite{wang2011surface}.  Furthermore, the 10-100nm length scale of quantum dots allows compact quantum computing architecture which can be mass-produced \cite{vandersypen2017interfacing}. However, this will make addressability of individual qubits more challenging.
Indeed, there have been multiple techniques employed in the literature in an attempt to address individual spins. Electron spins can be manipulated by coupling an applied ac electric field to the electron spin via spin-orbit coupling to achieve electrically driven spin resonance (EDSR) \cite{golovach2006electric}. Spins can then be driven locally via the gate electrodes \cite{tokura2006coherent} and magnetic field gradients from proximal micromagnets increase the coupling strength far above the intrinsic spin-orbit coupling and also lead to different resonant frequencies between neighboring qubits for additional distinguishability \cite{watson2018programmable}. This approach has proven quite successful with small numbers of qubits, although the on-chip microwave signals can lead to heating issues as the devices are scaled up \cite{vahapoglu2022coherent}. 

On the other hand, one can use an off-chip global microwave field to perform local rotation by electron spin resonance (ESR). In this case addressability can be achieved if each spin has a unique Larmor frequency. The Larmor frequency depends on the microscopic environment of the spin and there is a distribution of frequencies due to interface roughness and charge defects \cite{jones2018logical,ruskov2018electron,martinez2022variability}. When there are multiple qubits with similar Larmor frequencies, as is inevitable in any large array, addressability is compromised. The unwanted rotation of qubits near resonance with the target qubit is often referred to as crosstalk. There are protocols to avoid crosstalk by using gate voltages to locally shift Larmor frequencies in or out of resonance with the ESR field \cite{laucht2015electrically,seedhouse2021quantum, hansen2021pulse}, but this requires substantial tunability of the Larmor frequency, especially at high microwave power, to avoid off-resonant rotation. These issues were clearly discussed in a recent work by Cifuentes et al.~\cite{cifuentes2023bounds}, who stated, ``Strategies for qubit control need to be designed to circumvent this variability and tolerate the natural dispersion in qubit frequencies introduced by the oxide interface."

This paper provides a control strategy that can address a single qubit in an array of qubits using limited tunability of the Larmor frequency by leveraging the local controllability of the exchange coupling. In Sec.~\ref{theoretical model} we introduce a new pulse sequence, and in Sec.~\ref{error analysis} we show its performance versus the system size, finding that multi-qubit fidelity decreases only linearly with the number of qubits rather than exponentially.

\begin{figure}
    \centering
    \includegraphics[scale = 0.579]{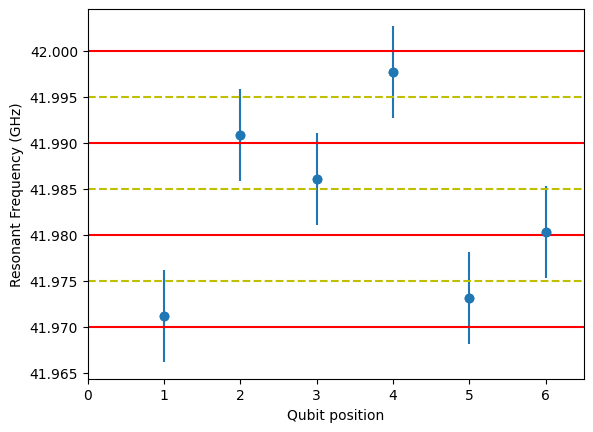}
    \caption{Example of resonant frequencies of spin qubits in a linear array. The red lines demarcate different frequency bins and the dashed lines indicate the central frequency of that particular frequency bin. In this example each qubit's frequency can be tuned up to 5 MHz in either direction (as shown by the vertical bars), so all qubits can be tuned to a central frequency.}
    \label{res-freq}
\end{figure}
\begin{figure}
    \centering
    \includegraphics[scale = 0.581]{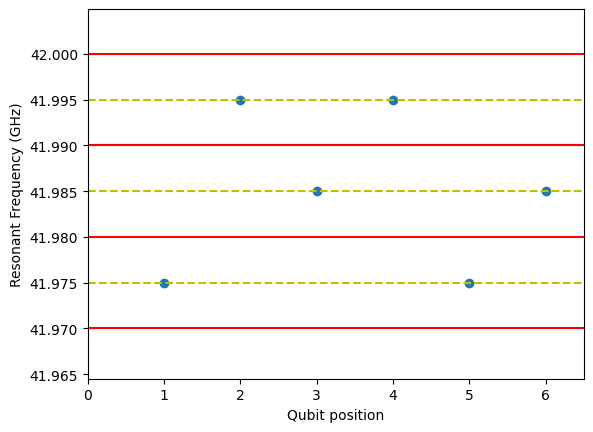}
    \caption{Example of resonant frequencies of spin qubits in a linear array after tuning to the central frequencies of the bins. Uniquely addressing a given target qubit now involves distinguishing it from the other qubits in the same bin using {\sc swap}s as well as managing a discrete set of off-resonance rotations.}
    \label{tuned res-freq}
\end{figure}

\section{Theoretical model} \label{theoretical model}
We introduce here a method to address a single spin in a linear chain with nearest neighbor exchange coupling. For our model, we consider an arbitrary number of spin qubits, each with a different resonant frequency. The system can be described by the Hamiltonian
\begin{equation}\label{eq:H}
    H = \sum_{i=1}^{N-1} J_{i} S_i \cdot S_{i+1} + \sum_{i=1}^N g_i \mu_B S_i \cdot B_i,
\end{equation}
where $J_{i}$ is the locally tunable exchange interaction between neighboring spins $S_i$ and $S_{i+1}$, and $B_i = (B_1 \cos \omega t, 0, B_0)$ is the magnetic field at the position of spin $S_i$ consisting of a homogeneous in-plane dc field $B_0$ and a perpendicular global microwave field of amplitude $B_1$. The resonant frequency of each qubit depends on its g-factor. 

The g-factor is weakly dependent on the spin-orbit coupling \cite{veldhorst2015spin} which in turn is affected by the random surface roughness and can be deliberately tuned by applying electric and/or magnetic fields \cite{rahman2009gate}. The Larmor frequency of a given qubit can thus be shifted by tuning one of the top gates that define that dot such that its $g-$factor is Stark shifted \cite{rahman2009gate,ruskov2018electron}, which allows for a variety of straightforward methods to address a single qubit with global control by shifting it in and out of resonance \cite{veldhorst2014addressable,veldhorst2017silicon,seedhouse2021quantum}.

Electrically controlled shifts on the order of $\pm 5$MHz at $B=1.4$T have been demonstrated \cite{veldhorst2014addressable,veldhorst2015two,huang2019fidelity,tanttu2019}, but this is not enough to neglect off-resonant effects when driving with Rabi frequencies $\sim 1$MHz. Additionally, schemes that park all idle qubits at resonance to take advantage of the enhanced robustness of the resulting dressed qubits \cite{seedhouse2021quantum} require enough tunability to encompass the entire spread of the qubits' frequency distribution, and the standard deviation of this distribution can be on the order of 60 MHz \cite{jones2018logical, cifuentes2023bounds}.

Fig.~\ref{res-freq} depicts an example of the resonant frequencies of spins in an array of six qubits. We group spins according to their resonant frequencies in frequency bins. The bin size is chosen to match the range of electrical tunability of the Larmor frequencies so that all qubits can be tuned to be at the center of a frequency bin. In this way the problem of addressability breaks into two separate parts: i) distinguishing the target qubit from other qubits in the same bin, which now all have identical frequencies, and ii) suppressing crosstalk between qubits in  different bins, which now have a discrete set of frequency differences. We will consider these two tasks in the following subsections.

\subsection{Pulse sequence to address one spin in a bin}\label{subsec:sequence}
Here, we propose a series of steps involving single qubit rotations and {\sc swap} operations to address a single target spin. The steps are enumerated below, starting with all exchange couplings turned off and no microwave field.
\begin{enumerate}
    \item Pulse the microwave to resonantly rotate the target qubit spin state around the $x$-axis. All the qubits in the same frequency bin as the target qubit also undergo a resonant rotation.
    \item {\sc swap} the target qubit with a qubit in a different bin which experiences negligible crosstalk with the target bin, by pulsing the exchange coupling links (usually it will suffice to pulse a single exchange link to {\sc swap} with a nearest neighbor, but if the nearest neighbors are all in the same bin as the target, one must sequentially {\sc swap} the state further along the array). The spin state of the target qubit is thus transferred to a separately addressable bin.
    \item Pulse the microwave at the new resonant frequency of the target state so as to rotate it about the $y$-axis. All the qubits in this frequency bin also undergo a resonant rotation.
    \item {\sc swap} the qubits again to return the spin states to their original locations.
    \item Pulse the microwave at resonance with the target qubit as in step 1 but so as to perform the inverse rotation.
    \item {\sc swap} the target qubit again with the same neighbor in step 2. 
    \item Pulse as in step 3 but so as to perform the inverse rotation.
    \item {\sc swap} the qubits back again as in step 4.
\end{enumerate}
At the end of this sequence only the target qubit acquires a net rotation, provided off-resonant rotations of qubits in one bin due to driving at the central frequency of another bin are negligible.
In other words, by using the local exchange control we can effectively dynamically toggle the resonant frequency of the target qubit between values beyond the range of the available in-situ tunability, although binning also provides substantially more power than this simple picture suggests, as Sec.~\ref{subsec:crosstalk} will make clear.

To make the logic more clear, the eight steps and the corresponding unitary operations are tabulated in Table \ref{tab:Table1} for the specific example of the g-factors taken in Figs.~\ref{res-freq}-\ref{tuned res-freq} and the target qubit being at position 1. The subscripts indicate the \emph{initial location} of each spin state for this example. Typically we envision the {\sc swap} being performed between the target and its nearest neighbor, but that doesn't need to be the case. The target qubit could be swapped with any qubit that is in a different frequency bin.

The ultimate rotation of the target qubit is 
\begin{equation}\label{eq:target rot}
    U = Y_{-\phi} X_{-\theta} Y_{\phi} X_{\theta}
\end{equation}
where $X_{\theta} \equiv \exp \left(-i\theta X/2\right)$ and $X$ is a Pauli operator. One can verify that setting $\phi = \theta$ produces the rotation
\begin{align}
    U &= Z_{\lambda} X_{\beta} Z_{\nu},\\
    \lambda &= -\frac{\pi}{4} - \arctan\left(\frac{\sin^2 \theta}{\sin^2\theta + 2\cos \theta}\right),\\
    \beta &= 2\arcsin\left(\sqrt{2}\sin^2 \frac{\theta}{2} \sin \theta\right),\\
    \nu &=  \frac{\pi}{4} - \arctan\left(\frac{\sin^2 \theta}{\sin^2\theta + 2\cos \theta}\right).
\end{align}
On the other hand, an arbitrary rotation can be formed by the Euler angle decomposition,
\begin{equation}\label{eq: SU2 rotation}
    U(\alpha, \beta, \gamma) = Z_{\alpha} X_{\beta} Z_{\gamma},
\end{equation}
where $\alpha$, $\beta$, and $\gamma$ are the Euler angles. Thus, our sequence, supplemented by additional $Z$ rotations which can be executed virtually (i.e., by instantaneous changes in the phase of the rotating reference frame) \cite{mckay2017efficient}, is capable of producing any rotation such that $|\beta|<2\arcsin\left(\frac{3\sqrt{3}}{4\sqrt{2}}\right) \approx 0.74\pi$. Unitaries requiring larger $\beta$ can be formed from two $X_{\pi/2}$ rotations interleaved with virtual $Z$ rotations. Thus this sequence is universal for single-qubit gates.

\begin{table}
\caption{Evolution of the spins of Figs.~\ref{res-freq}-\ref{tuned res-freq} under the eight-step sequence.}
\begin{tabular}{ |c||c||c|  }
 \hline
1 & Bin 1 rotation ($X_{\theta}$) & $X_{\theta,1} I_2 I_3 I_4 X_{\theta,5} I_6 $  \\
 \hline
2 & {\sc swap} qubits 1 and 2 & $ I_2 X_{\theta,1}  I_3 I_4 X_{\theta,5} I_6$ \\
\hline
3 & Bin 3 rotation ($Y_{\phi}$)  & $ I_2 (Y_{\phi} X_{\theta})_1 I_3 Y_{\phi, 4} X_{\theta,5} I_6
$ \\
\hline
4 & {{\sc swap} qubits 1 and 2}  & $ (Y_{\phi} X_{\theta})_1 I_2 I_3 Y_{\phi, 4} X_{\theta,5} I_6 $ \\
\hline
5 & Bin 1 rotation ($X_{-\theta}$)  & $(X_{-\theta} Y_{\phi} X_{\theta})_1 I_2 I_3 Y_{\phi, 4} I_5 I_6 $ \\
\hline
6 & {{\sc swap} qubits 1 and 2} & $ I_2 (X_{-\theta} Y_{\phi} X_{\theta})_1 I_3 Y_{\phi, 4} I_5 I_6$ \\
\hline
7 & Bin 3 rotation ($Y_{-\phi}$) & $I_2 (Y_{-\phi}X_{-\theta} Y_{\phi} X_{\theta})_1 I_3 I_4  I_5 I_6$ \\
\hline
8 & {{\sc swap} qubits 1 and 2}  & $(Y_{-\phi}X_{-\theta} Y_{\phi} X_{\theta})_1 I_2 I_3 I_4  I_5 I_6$ \\
\hline
\end{tabular}
\label{tab:Table1}
\end{table}

We now briefly note how the {\sc swap} operations are performed. In the absence of driving, the Hamiltonian in Eq.~\eqref{eq:H} for a given pair of neighboring qubits can be written as
\begin{multline}\label{eq:Hswap}
    H = \frac{J}{4} \left(XX+YY+ZZ\right) 
    + \frac{\Delta E_z}{2} (IZ-ZI) 
    \\
    + \overline{E_z} (IZ+ZI),
\end{multline}
where $J = J_{i}$, $\Delta E_z = \mu_B B_0 (g_{i+1} - g_i)$, and $\overline{E_z} = \mu_B B_0 (g_{i+1} + g_i)/2$. Note that the Hamiltonian can be split into two mutually commuting terms and belongs to an embedding $\mathfrak{su}(2)\oplus\mathfrak{u}(1) \subset \mathfrak{su}(4)$. The $\mathfrak{u}(1)$ part of the Hamiltonian is 
\begin{equation}\label{Hu1}
    H_{\mathfrak{u(1)}} = \frac{J}{4} ZZ + \overline{E_z} (IZ+ZI)
\end{equation}
and the $\mathfrak{su}(2)$ part of the Hamiltonian can be written as
\begin{equation}\label{Hsu2}
    H_{\mathfrak{su}(2)} = \frac{J}{2} \tilde{Z} + \Delta E_z \tilde{X},
\end{equation}
where $\tilde{Z} = (XX+YY)/2$ and $\tilde{X} = (IZ-ZI)/2$.

A gate equivalent to {\sc swap} (up to local $z$ rotations) is obtained by performing a $\pm\pi/2$ rotation in the effective $SU(2)$ about $\tilde{Z}$. If one can access high enough exchange coupling that the effect of the nonzero $\Delta E_z$ is negligible, then the {\sc swap} can be done in time $T_{\text{\sc swap}} = \pi/2J$. Generally, though, one may not be able to access high enough exchange for this direct implementation, but one can compose the rotation with an exchange pulse sequence \cite{ramon2011electrically}
\begin{equation}\label{Rz}
    R(\hat{z}, \alpha) = R(\sin \gamma \hat{x}+ \cos \gamma \hat{z}, \chi) R(\hat{x}, \varphi) R(\sin \gamma \hat{x}+ \cos \gamma \hat{z}, \chi)
\end{equation}
where $R(\hat{n}, \chi) = \exp{(-i \frac{\chi}{2} \hat{n}\cdot \vec{\sigma})}$ is a rotation about the axis $\hat{n}$ by an angle $\chi$. 
The solutions for $\chi$ and $\varphi$ in terms of $\gamma$ and $\alpha$ are (rearranging expressions from Ref.~\cite{throckmorton2017fast} and noting that $J\geq 0$ implies $-\pi/2\leq \gamma \leq \pi/2$)
\begin{equation}\label{alpha}
    \varphi = -2 \arcsin\left[ \tan \gamma \sin \left( \frac{\alpha}{2} \right) \right],
\end{equation}
\begin{equation}\label{chi}
    \chi = \sgn \alpha \arccos \left[1 - \frac{1 - \sqrt{\cos^2 \left( \frac{\alpha}{2} \right)- \frac{1}{4} \sin^2 \alpha \tan^2 \gamma}}
    {\cos^2\left( \frac{\alpha}{2}\right) + \sin^2\left( \frac{\alpha}{2}\right) \cos^2 \gamma} \right].
\end{equation}
In our application, the largest accessible value of exchange determines the available $\gamma = \arctan(2\Delta E_z/J)$ and the middle segment is performed by turning the exchange off for a time. Note that, physically, $\chi$ must be positive and $\varphi$ must have the same sign as $\Delta E_z$, which can be ensured by adding or subtracting $2\pi$ to the rhs of Eqs.~\eqref{alpha}-\eqref{chi} as needed. Also note that Eqs.~\eqref{alpha}-\eqref{chi} are real only if $|2\Delta E_z/J| \leq |\csc (\alpha/2)|$, so for any neighboring pair of qubits where that is not the case for $\alpha = \pm \pi/2$, one can instead use $\alpha = \pi/2n$ with the integer $n$ chosen to ensure Eq.~\eqref{alpha} remains real, and the composite rotation of Eq.~\eqref{Rz} can be repeated $n$ times to produce the {\sc swap}.
The total duration of the {\sc swap} gate is thus
\begin{equation}\label{t_swap}
    T_{\text{\sc SWAP}} = n\frac{2 \chi \cos \gamma + \varphi \cot \gamma}{J}, \quad n= \lceil \frac{\pi}{4|\arcsin (J/2\Delta E_z)|} \rceil
\end{equation}
which ranges from $(3.5\pi+\sqrt{2})/J$ when $\Delta E_z\ll J$ down to $\pi^2/2J$ when $\Delta E_z \gg J$.

Alternatively, specific to the case $J>2\Delta Ez$, there is also a simpler and faster composition that does not turn exchange all the way to zero in the middle segment \cite{Zhang2017}.

\subsection{Suppressing interbin crosstalk}\label{subsec:crosstalk}
Consider a constant amplitude global pulse intended to perform a resonant $X_{\theta}$ rotation with exchange coupling turned off, as in step 1 of the sequence above. ``Idle," off-resonant qubits have a Hamiltonian of the form
\begin{equation}\label{eq:UncoupledH}
    H = g \mu_B S \cdot B = \hbar \left(
\begin{array}{c c}
\omega_0/2 & \Omega \cos\omega t\\
\Omega \cos\omega t & -\omega_0/2
\end{array}\right)\;,
\end{equation}
where $\omega_0$ is the resonant frequency, $\Omega$ is the driving strength, and $\omega$ is the driving frequency. In the rotating frame $H_r = R^{\dagger} H R - i R^{\dagger}\dot{R}$, with $R= \exp(-i \omega t/2 \,Z)$, and rotating wave approximation (RWA) gives
\begin{equation}\label{eq:RWA-H}
    H_r = \frac{\hbar\Omega}{2} X + \frac{\hbar m \delta}{2} Z
\end{equation}
and an evolution operator in the rotating frame after time $T$
\begin{equation}\label{ORR-unitary}
    U = e^{-iT (\Omega X + m \delta Z)/2},
\end{equation}
where $m \delta = \omega_0-\omega$ is the detuning from resonance with $m$ as the integer difference between the index of the resonant bin and the off-resonant bin and $\delta$ is the bin width.

While $z$-rotations are not worrisome since they can be absorbed into the local rotating frame (i.e., compensated by virtual $z$-rotations \cite{mckay2017efficient}), we wish to avoid rotation along any other axis resulting from the off-resonant driving. There is clearly no $y$-component to the off-resonant rotation of Eq.~\eqref{ORR-unitary}, but there is also no $x$-component if one chooses 
\begin{equation}
    \sqrt{\Omega^2 + (m \delta)^2}T = 2n\pi, \quad n \in \mathcal{Z}.
\end{equation}
Substituting in that one must also have 
\begin{equation}\label{eq:T}
    \Omega T = \theta    
\end{equation}
for the driving to produce the intended resonant rotation, one obtains a condition similar to the synchronization condition found in \cite{heinz2021crosstalk},
\begin{equation}\label{Omega-expression}
    \Omega = \pm\frac{m \delta \theta}{ \sqrt{ \left(2n \pi\right)^2 - \theta^2}}.
\end{equation}
The same condition also holds for resonant $y$-rotations to have negligible off-resonant effects. Note that to ensure real values of $\Omega$ the choice of integer $n$ must satisfy
\begin{equation}\label{n-m bound}
    n > \frac{\theta}{2\pi}.
\end{equation}
Thus, by careful choice of the driving amplitude (and time), one can completely suppress off-resonant errors at a given detuning. However, instead of aiming for complete error suppression in a given bin, we would like to minimize the overall infidelity arising from a large set of bins, most of which will be many bins away from the target bin. Thus we consider the limit of large $m$. Note that by choosing the free integer to be $n=\ell m$, the optimal driving strength of Eq.~\eqref{Omega-expression} becomes independent of bin index for large $m$,
\begin{equation}\label{eq: refined omega}
    \Omega = \frac{\delta \theta}{2 \ell \pi}, \quad \ell \in \mathcal{Z}.
\end{equation}
In fact, although this is obtained for the large $m$ limit, it is nearly optimal even for the nearest bin, $m=1$, since the denominator of Eq.~\eqref{Omega-expression}, $\sqrt{(2 \ell \pi)^2-\theta^2} \approx 2 \ell \pi$ for $\theta \ll 2\ell \pi$. So, by driving with one of these particular strengths we suppress off-resonant effects across all bins. Larger values of $\ell$ produce a better approximation at the cost of longer pulse times.

\section{Sequence performance} \label{error analysis}
Now that we have established the pulse sequences to address the target qubit and the optimal parameters for suppressing off-resonant rotations, we analyze the effects of residual off-resonant rotations on the entire sequence. We will not include the effect of local pulse miscalibration errors in our analysis because those errors are independent of the adressability errors and their effect on the multi-qubit fidelity is comparatively small. Furthermore, those errors can in principle be eliminated beforehand by local tune-up procedures. Neither do we consider the effect of charge noise, since again this is a separate issue whose effect on the multi-qubit fidelity is small in comparison when the total sequence time is much less than the dephasing time $T_2^{\ast}$. In principle, even this restriction could be relaxed by using dynamically correction via broadband pulses that can simultaneously invert all the spins or by more sophisticated sequences or pulse shaping \cite{Kanaar2022}. For the present though, since a typical dephasing time is $T_2^{\ast} \sim 100 \mu$s, it is sufficient to restrict ourselves to sequences of length $\sim 10 \mu$s.

The total time for the sequence is
\begin{equation}\label{total time }
    T_{total} = 4 \times T + 4 \times T_{\text{\sc swap}},
\end{equation}
where the local rotation step time is given by Eqs.~\eqref{eq:T} and \eqref{eq: refined omega} as 
\begin{equation}
    T = 2\ell\pi / \delta.
\end{equation} 

The multi-qubit fidelity for one local rotation in the sequence (steps 1, 3, 5, or 7) at the frequency of target bin $t$ is given by 
\begin{equation}\label{total fidelity simplified}
    \mathcal{F}_{loc} (t) = \prod_{j \neq 0} |\frac{1}{2}\text{Tr}(U_j)^{N_{t+j}}|^2,
\end{equation}
where $U_j = e^{-iT (\Omega X + j\delta Z)/2}$, and $N_{t+j}$ is the number of qubits in the $(t+j)$-th bin. 

The entire sequence for a particular configuration $\vec{N} = (N_1, N_2, \dots)$ has a fidelity whose lower bound is the product of the fidelities of each step of the sequence. Assuming the fidelity of the nearest neighbor {\sc swap} gates, $\mathcal{F}_{\text{\sc swap}}$, are independent of the qubit frequencies, averaging over all possible target qubits in the configuration gives a sequence fidelity of
\begin{equation}\label{eq:seq fid}
    \mathcal{F}_{seq} (\vec{N}) \geq \mathcal{F}_{\text{\sc swap}}^4 \sum_{t}\sum_{k\neq t} \frac{N_t}{N} \frac{N_k}{N-N_t} \mathcal{F}_{loc}(t)^2 \mathcal{F}_{loc}(k)^2,
\end{equation}
since when randomly selecting a target qubit, the probability of it being in bin $t$ is $N_t/N$ and the probability that the nearest neighbor qubit (excluding qubits in the target bin) is in bin $k$ is $N_k/(N-N_t)$. We set $\mathcal{F}_{\text{\sc swap}}=1$ in the numerics below in order to focus on addressability errors.

To calculate the average total fidelity of the $N$-qubit system we randomly sample a large number qubit configurations. Each configuration is generated by randomly drawing $N$ times from a normal distribution of mean $\omega_0$ and standard deviation $\sigma$. We calculate the weighted average of Eq.~\eqref{eq:seq fid} over these configurations,
\begin{equation}\label{eq: F_avg for the whole pulse}
    \mathcal{F}_{avg} = \left(\sum_{\vec{N}} p (\vec{N})\right)^{-1}
     \sum_{\vec{N}} p(\vec{N}) \mathcal{F}_{seq}(\vec{N}).
\end{equation}
Here the probability of a given configuration is
\begin{equation}\label{eq: p_n}
    p(\vec{N}) =  \prod_{j=-\infty}^{\infty} p_j^{N_j} \binom{N-\sum_{k=-\infty}^{j-1} N_k}{N_j}
\end{equation}
where
\begin{equation}\label{eq: p_t}
    p_j = \int_{\omega_0+(j-1/2)\delta}^{\omega_0+(j+1/2)\delta} \frac{1}{\sqrt{2 \pi \sigma^2}} e^{- (\omega-\omega_0)^2/2 \sigma^2} d\omega,
\end{equation}
is the probability of a given qubit being in bin $j$.

For specificity, we numerically consider the case of where one wishes to apply a local $X_{\pi/2}$ rotation, although it is easy to generalize to an arbitrary rotation. For this case, $\phi=\theta=\pi/2$ in Eq.~\eqref{eq:target rot}. We take an experimentally plausible bin width $\delta = 10$MHz and a distribution of width $\sigma = 60$ MHz for a magnetic field oriented along the [110] lattice direction \cite{cifuentes2023bounds}. The rms average of $\Delta E_z$ is $\sqrt{2}\sigma \approx 85$MHz and taking a maximum exchange coupling of $J \sim 50$MHz, the average $T_{\text{\sc swap}} \sim 0.1 \mu$s from Eq.~\eqref{t_swap}. Then the optimal driving strength for all local rotation steps is given by Eq.~\eqref{eq: refined omega}, choosing $\ell = 4$ to keep the total time to $\sim 10\mu$s, as $\Omega = 0.625$ MHz. In that case there is suppressed off-resonant rotation of qubits in neighboring bins such that they each undergo an identity operation with fidelities of 
\begin{equation}
    |\frac{1}{2}\text{Tr}(U_{\pm 1})|^2 = 0.9994, \quad |\frac{1}{2}\text{Tr}(U_{\pm 2})|^2 = 0.99985,
\end{equation}
and so on, with higher fidelity in all further bins. As we increased the number of configurations included in our numerical computation of Eq.~\eqref{eq: F_avg for the whole pulse} from $10^{3}$ to $10^5$, the resulting values of $\mathcal{F}_{avg}$ only changed by $\sim 10^{-5}$, suggesting our numerics are well converged.

\begin{figure}
    \centering
    \includegraphics[scale = 0.133]{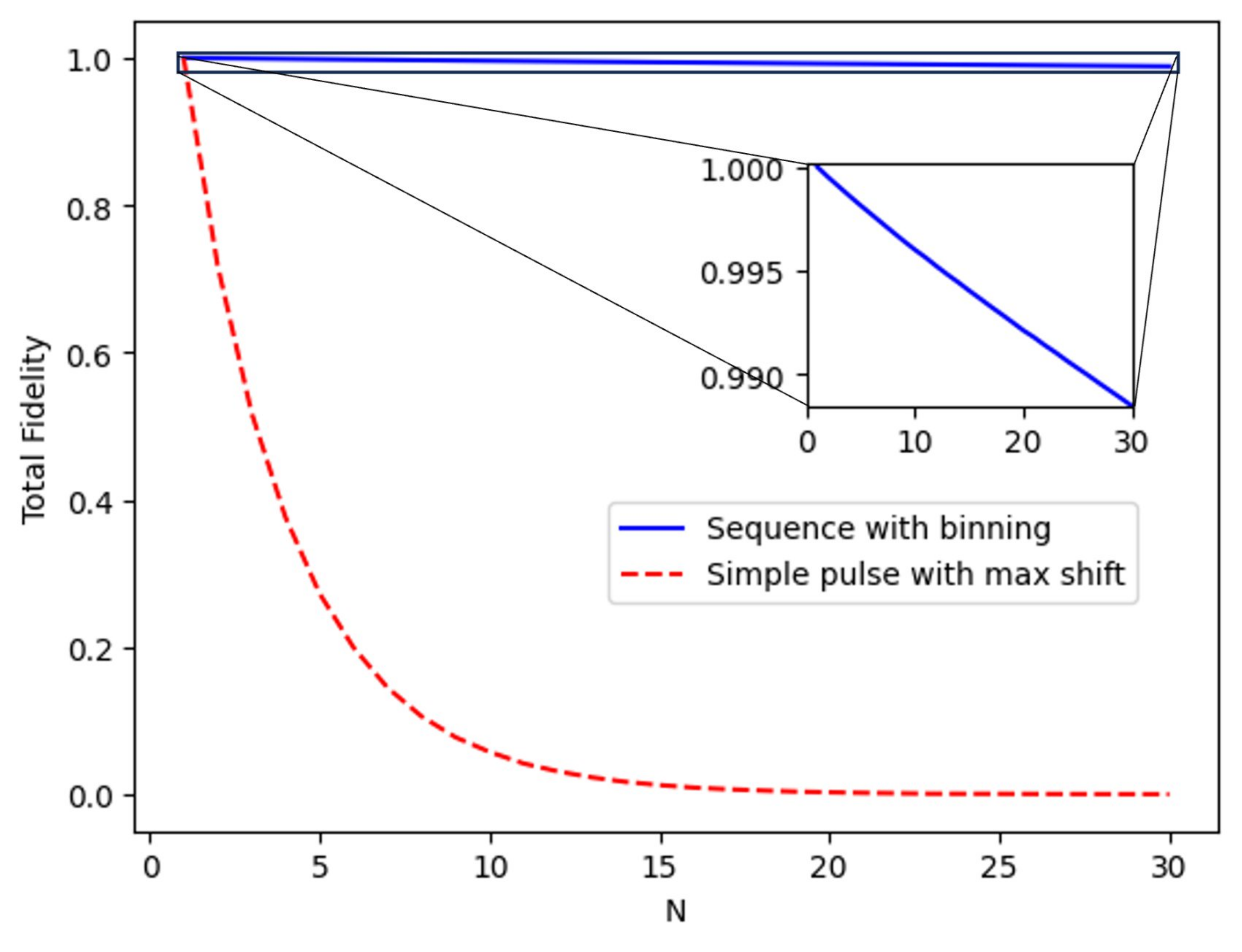}
    \caption{Average N-qubit fidelity vs the number of qubits for a single-qubit $X_{\pi/2}$ rotation in a qubit array whose Larmor frequencies are normally distributed with standard deviation $\sigma = 60$ MHz and individually tunable by $\pm 5$ MHz. Both gates are of duration $\sim 10 \mu$s.}
        \label{tuned vs untuned fidelities}
\end{figure}
The results are plotted in Fig.~\ref{tuned vs untuned fidelities}, comparing the total fidelity of the proposed pulse sequence on the qubits after they are tuned to the central frequency of each frequency bin to the total fidelity in the naive case of a single, simple pulse at the resonant frequency of the target qubit after each idle qubit frequency is shifted as far away as possible from the target given the limited tunability of $\pm \delta/2$. Since slower pulses are more frequency selective, we allowed the simple pulse to use the same time as the full sequence to make a fair comparison, so the driving strength of the simple pulse was reduced to $\Omega_{simple} = \pi/2T_{total} \approx 0.157$ MHz. The significant advantage of our method is evident from Fig.~\ref{tuned vs untuned fidelities} in that the decay of multi-qubit fidelity with the number of qubits is exponential for the simple pulse and only linear for the sequence.

\section{Conclusion} \label{conclusion}
In this paper, we have demonstrated an effective method for addressing a specific qubit within a large array of qubits using a combination of global microwave pulses and composite {\sc swap} gates. The multi-qubit fidelity of a single-qubit gate decays only linearly in the number of qubits when using our approach, compared to exponentially with the naive approach.
For realistic parameter values, we numerically obtain a fidelity of over 99\% for a system of as large as 25 qubits.

One could also use this approach while simultaneously addresssing selected multiple qubits by carrying out the sequence in parallel with different targets, provided that the frequencies of the two bins driven for each target (the original bin and the bin after the {\sc swap}) only cause non-negligible off-resonant rotations on a certain set of bins and the sets for the different targets are disjoint.

\begin{acknowledgments}
The authors acknowledge useful discussions with David Kanaar. This research was sponsored by the Army Research Office (ARO), and was accomplished under Grant Numbers W911NF-17-1-0287 and W911NF-23-1-0115.
\end{acknowledgments}

\bibliography{apssamp}

\end{document}